\documentclass{INTERSPEECH2023}
\usepackage{multirow}
\interspeechcameraready

\title{Study of GANs for Noisy Speech Simulation from Clean Speech}
\name{Leander Melroy Maben $^1$, 
Zixun Guo$^2$, 
Chen Chen$^2$, 
Utkarsh Chudiwal$^3$ 
and
Chng Eng Siong $^2$}
\address{
  $^1$Manipal Institute of Technology, Manipal, India
  $^2$Nanyang Technological University, Singapore
  $^3$Indian Institute of Technology, Indore, India
  }
\email{leander.maben@gmail.com, zguo008@ntu.edu.sg, chen1436@e.ntu.edu.sg, utkarsh.chudiwal@gmail.com, aseschng@ntu.edu.sg}

\begin{document}

\maketitle
 
\begin{abstract}
The performance of speech processing models trained on clean speech drops significantly in noisy conditions. Training with noisy datasets alleviates the problem, but procuring such datasets is not always feasible. Noisy speech simulation models that generate noisy speech from clean speech help remedy this issue. In our work, we study the ability of Generative Adversarial Networks (GANs) to simulate a variety of noises. Noise from the Ultra-High-Frequency/Very-High-Frequency (UHF/VHF), additive stationary and non-stationary, and codec distortion categories are studied. We propose four GANs, including the non-parallel translators, SpeechAttentionGAN, SimuGAN, and MaskCycleGAN-Augment, and the parallel translator, Speech2Speech-Augment. We achieved improvements of 55.8\%, 28.9\%, and 22.8\% in terms of Multi-Scale Spectral Loss (MSSL) as compared to the baseline for the RATS, TIMIT-Cabin, and TIMIT-Helicopter datasets, respectively, after training on small datasets of about 3 minutes.
\end{abstract}

\section{Introduction}

Audio processing has wide-ranging applications in a multitude of tasks. Advancements in deep learning have further enabled the large-scale use of audio processing in tasks like Automatic Speech Recognition and Speech Enhancement \cite{eesen,deepspeech,https://doi.org/10.48550/arxiv.1904.12069}. The major challenge, however, is that the noisy environments that these models are exposed to in the real world severely impact their performance, making exposure to noise during training essential for good performance. Several works \cite{badrinath2022automatic, pellegrini2018airbus, godfrey1992switchboard} use corpora of noises obtained from real-world settings. However, gathering noise datasets with sufficient variations and volume is not always economical or feasible.

Artificially generating noisy data is one viable option to remedy this issue. Several traditional models exist for audio augmentation, including parallel model matching \cite{gales1995model}, HMM adaptation using vector Taylor series \cite{acero2000hmm}, and stochastic matching \cite{sankar1995robust}. There have also been works that directly add short static noise clips to clean audios for augmentation \cite{ma2019improving}. Obtaining significant durations of real-world noise, however, is often infeasible. Works like \cite{ferras2016large}, and \cite{9023257} use codecs to simulate distortion. Most of these traditional augmentation methods are based on using short clips of noise or pre-defined processing techniques. Using short noise clips entails the drawback of exposing the model to low levels of variation and hence lowers the model's generalizability to the broader variation levels present in the real world. Furthermore, if the augmentation techniques are predefined and not learned, they cannot adapt to different noise types.

Techniques involving deep learning or differentiable digital signal processing (DDSP) \cite{engel2020ddsp, dent-ddsp} where model parameters are learned show potential in being more adaptable to different noise domains. Among deep learning techniques, GANs show great promise in boosting model performance in noisy conditions due to their ability to simulate data realistically. GANs have the ability to perform both parallel and non-parallel translations from one domain to another. Parallel translation involves  training the translation model with pairs of corresponding data points from both domains in a supervised fashion. On the other hand, non-parallel translation does not require training with input-target pairs, and hence, training is done in an unsupervised manner.

SimuGAN \cite{9747755} has shown promise in simulating noise by learning noise statistics and has been shown to boost the performance of Automatic Speech Recognition (ASR). However, it has only been tested on a single type of noise and does not explore the usability of this GAN-based technique with other noise categories. DENT-DDSP \cite{dent-ddsp} uses differentiable digital signal processing for parallel translation from the clean to the noisy domain in a very data-efficient manner. However, obtaining parallel data is difficult, and this may make the method unusable in applications where only non-parallel data is available. 

There has been limited exploration of the ability of GANs to simulate noises of different types under conditions of low data volume constraints. In our work, we aim to explore the ability of four types of GANs to simulate four types of noises. We use SpeechAttentionGAN, MaskCycleGAN-Augment, and SimuGAN for non-parallel translation and Speech2Speech-Augment for parallel translation. SpeechAttentionGAN and MaskCycleGAN-Augment belong to the cycleGAN \cite{https://doi.org/10.48550/arxiv.1703.10593} family and use the 'Filling in Frames' technique during training. In addition, SpeechAttentionGAN uses attention masks for translation. SimuGAN uses the PatchNCE \cite{patchNCE} loss for content preservation. Speech2Speech-Augment uses supervised learning with L1 loss to learn the translation. Furthermore, we use the RATS dataset, cabin and helicopter noise from the TIMIT dataset, and codec2 to represent the noise categories of UHF/VHF, additive stationary, additive non-stationary, and codec-based distortion, respectively. We also compare the performance of noise simulation with GANs against a baseline. 

Our work proves that GANs can be used to simulate noisy speech through experimental evidence. All four types of GANs show superior simulation quality in terms of LSD and MSSL when compared against the baseline. Unlike the traditional models, our proposed models learn to simulate the noise and are not based on a fixed processing algorithm or using real-world noise clips. It thus provides great flexibility to be used for various kinds of noisy environments, as demonstrated by the experiments with multiple noise types. We also propose three GANs for non-parallel translation, which overcomes the drawbacks of learning-based techniques like DENT-DDSP \cite{dent-ddsp} that cannot be used without parallel data. Furthermore, our experiments were conducted using only about 3 minutes of clean and noisy clips each, demonstrating a way to effectively simulate noise even when the volume of data available is extremely low. Moreover, while traditional and learning-based methods involving DDSP cannot be used to simulate dynamic noise, we prove experimentally that GANs can successfully perform this task.  Finally, the detailed experimental results provide researchers and developers of audio augmentation models with a comprehensive framework to use the correct type of GAN for the right kind of noise simulation task. 

\section{Methodology}

\subsection{Overview of the Audio Processing Pipeline}
\begin{figure}
    \centering
    \includegraphics[width=0.3\textwidth]{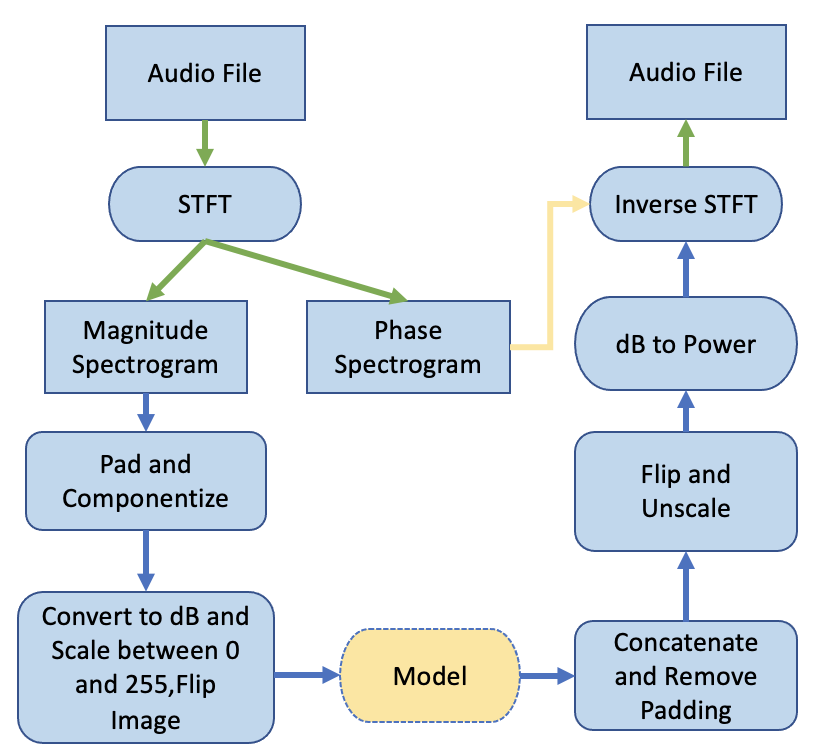}
    \caption{Overview of the Audio Processing Pipeline. The trainable model is depicted by the yellow component. The flow of the magnitude spectrogram is depicted by blue arrows, the flow of the phase spectrogram is depicted by the yellow arrows, and the combined information flow is depicted by green arrows.}
    \label{overview}
\end{figure}

Figure \ref{overview} gives an overview of the prominent steps in the pipeline that converts a clean audio file to a noisy audio file. These steps are as performed as follows. The loudness of the audio clip is first normalized to a fixed value. This is followed by Short Time Fourier Transform (STFT) to get a magnitude and phase spectrogram. Further processing is only done on the magnitude spectrogram. The phase spectrogram will be recycled and used in inverse STFT. The magnitude spectrogram goes through padding and componentization, which will be explained in detail in the following subsection. Each component then undergoes several processing steps, including conversion to decibels and scaling between 0 to 255. The magnitude spectrogram initially shows the lower frequencies on top and higher frequencies on the bottom. We flip it vertically to reverse this order. The components are then processed by the model to give us the generated components. The generated components are concatenated, and any padding added during componentization is removed. The magnitude spectrogram thus obtained is flipped, appropriately scaled, and converted to power from decibels. Inverse STFT is now performed using this generated magnitude spectrogram and the original phase spectrogram (from the clean audio) to get the target noisy audio clip.

\subsection{Componentization}

The lengths of spectrograms are variable and depend on the lengths of the audio clips. Longer audio clips would have longer spectrograms, while shorter audio clips would have shorter ones along the horizontal axis. The height of the spectrogram depends on the window length (after padding) parameter of the STFT and would remain constant as this parameter is kept constant for all audio samples. During componentization, the spectrogram is broken into components of constant length, and the last component is padded with the required number of frames from the start of the spectrogram if it is not long enough. This ensures that the model inputs are kept at constant dimensions and also helps a single audio clip to serve as multiple training points if it is long enough.

\subsection{GANS}

\subsubsection{SpeechAttentionGAN}

\begin{figure}
    \centering
    \includegraphics[scale=0.26]{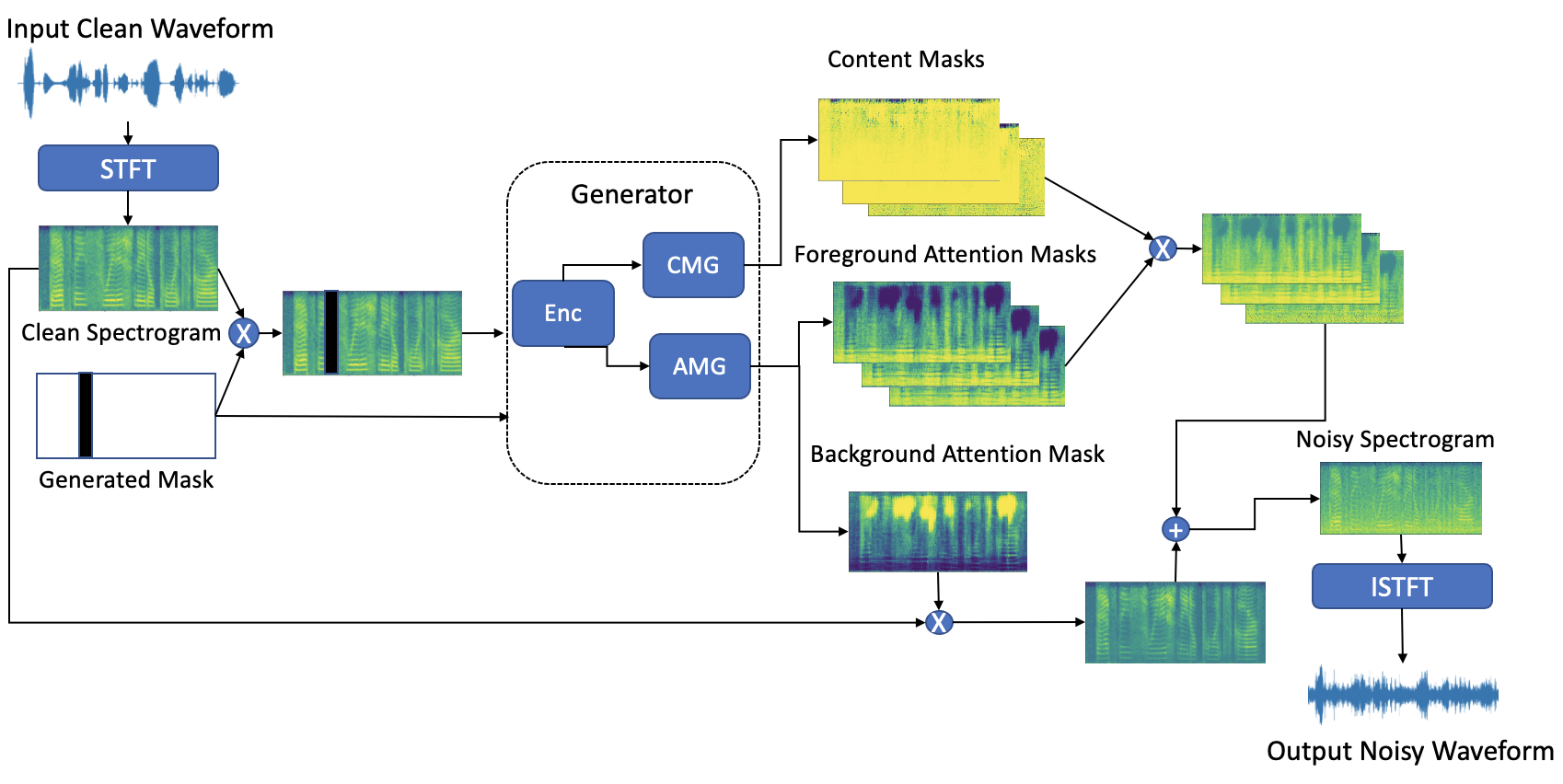}
    \caption{SpeechAttentionGAN. The figure depicts the overall flow of data in one direction in the SpeechAttentionGAN pipeline with Filling in Frames. The generator consists of three parts - the Encoder (Enc), the Attention Mask Generator(AMG), and the Content Mask Generator(CMG). In this figure, spectrogram refers to magnitude spectrogram, STFT refers to Short-Time Fourier Transform, and ISTFT refers to Inverse Short-Time Fourier Transform.}
    \label{fig:attention}
\end{figure}

SpeechAttentionGAN uses the base model architecture from AttentionGAN \cite{tang2021attentiongan, tang2019attention}, which was designed to be used for the non-parallel translation of images in a data-efficient manner. Information flow in SpeechAttentionGAN is depicted in figure \ref{fig:attention}.

 In the proposed model, we introduce two novel modifications on top of the base model to help the model learn spectrogram structures better and to remedy the statistical averaging caused by the cycle consistency loss \cite{https://doi.org/10.48550/arxiv.1703.10593}. They are the addition of an input mask to implement the 'Filling in Frames' (FIF) approach in model training and a second adversarial loss, respectively.

To implement FIF, we generate a mask of zeros and ones by which the input spectrogram is then multiplied. The mask is generated in such a way that all frequencies for a random range of time frames of the spectrogram are removed during training. The masked spectrogram and the mask are given as inputs to the model. By training in this fashion, the model is forced to predict the missing content from the surrounding information and thus learn the structure of the spectrogram implicitly. During testing, the entire mask is set to one, and no information loss takes place.

The second adversarial loss \cite{9414851} is very similar in structure to the general adversarial loss \cite{https://doi.org/10.48550/arxiv.1406.2661} used in GANs. The difference lies in the fact that instead of using the real input and the output generated by one generator, the second adversarial loss uses real inputs and cycled outputs (outputs that have been cycled through both generators back to the original domain). The second adversarial loss for the translation from domain A to domain B and back to domain A (A$\rightarrow$B$\rightarrow$A) is described in equation \ref{eq:adv2}. The second adversarial loss for the direction B$\rightarrow$A$\rightarrow$B is computed in a similar manner. In order to compute the second adversarial loss, we use two additional discriminators with identical architecture to the original two. We use one discriminator each to compute this loss in the two directions.

\begin{equation}
    {\scriptscriptstyle L_{adv2}^{A->B->A}=E_{a \sim A}[D'_{A}(a)]+E_{a \sim A}[D'_{A}(G_{B->A}(G_{A->B}(a)))]}
    \label{eq:adv2}
\end{equation}

In equation \ref{eq:adv2}, $G_{A->B}$ represents the generator that translates from domain A to domain B, $G_{B->A}$ represents the generator that translates from domain B to domain A, $D'_{A}$ represents the discriminator used to compute this loss in the direction A$\rightarrow$B$\rightarrow$A, and $a$ represents a data point from domain A. 

\subsubsection{MaskCycleGAN-Augment}

The MaskCycleGAN-Augment model uses the base architecture and framework followed by MaskCycleGAN-VC \cite{9414851}. MaskCycleGAN-VC uses MelGAN \cite{NEURIPS2019_6804c9bc} for the conversion between audio files and spectrograms.  This approach, however, does not work well for our use case of converting clean speech to noisy speech. In particular, upon conversion of a noisy spectrogram to an audio clip, MelGAN performs poorly and outputs unintelligible clips. As a remedy to this issue, MaskCycleGAN-Augment uses the audio processing pipeline described in subsection 2.1.

\subsubsection{SimuGAN}

SimuGAN \cite{9747755} tackles the problem of non-parallel translation and does not use cycle consistency loss \cite{https://doi.org/10.48550/arxiv.1703.10593}. Instead, it promotes content preservation by encouraging the model to generate outputs such that the corresponding patches in the input and output spectrograms have more similar latent representations than the non-corresponding patches using a patchNCE loss \cite{patchNCE}.

\subsubsection{Speech2Speech-Augment}

Speech2Speech-Augment uses the base architecture of the Pix2Pix \cite{8100115} model, which was designed for parallel image translation. It uses the adversarial loss \cite{https://doi.org/10.48550/arxiv.1406.2661} along with an L1 loss to encourage the model to produce outputs close to the expected targets.

\section{Experiments and Results}

\begin{figure*}
    \centering
    \includegraphics[width=0.9\textwidth]{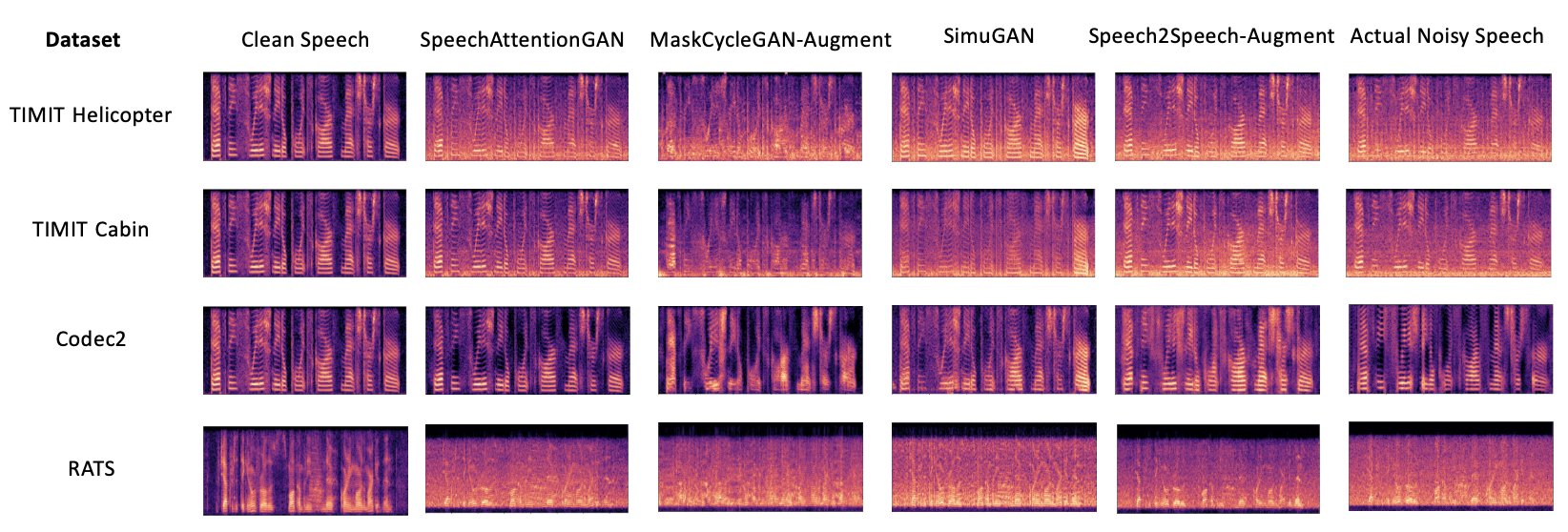}
    \caption{Generated Samples. This figure shows the noisy spectrogram samples generated by the four models for each of the four noise types. It also shows the clean and actual noisy spectrogram samples from the four datasets.}
    \label{fig:samples}
\end{figure*}
\subsection{Datasets}

In order to study the potency and performance of GANs in simulating a wide range of noises, we select noises from 4 different categories. These include Ultra-High-Frequency/Very-High-Frequency (UHF/VHF) noise represented by the RATS dataset \cite{upennRATSSpeech}, additive stationary and non-stationary noises represented by the cabin and helicopter noises from the TIMIT dataset \cite{upennTIMITAcousticPhonetic}, respectively, and codec distortion generated using codec2 \cite{rowetelCodecx2013}. The RATS dataset consists of clean and noisy speech. The TIMIT dataset, however, consists of clean clips along with clips of pure noise. We use the method described in \cite{NEURIPS2021_a5c7b30f} to generate noisy speech clips using clean clips and noise clips. Similarly, we use the clean clips from the TIMIT dataset and pass them through the codec to get noisy speech clips with codec distortion. All datasets used are in the English language.

We prepare both parallel and non-parallel training sets for each type of noise. The training sets for all types of noises include approximately 3 minutes each of noisy and clean audio clips. The validation and test sets, respectively, contain approximately 100 and 200 seconds each of clean and noisy audio clips.

Furthermore, for the TIMIT stationary and non-stationary datasets, as well as the codec2 dataset, we ensured that there were no common speakers between any two of the three splits.

\subsection{Experimental Setup}
During training, model checkpoints are saved once every fixed number of epochs. In the validation phase, the validation set is used to evaluate model performance at different checkpoints. The model checkpoint with the best performance on the validation set is picked and used to generate simulations on the test set. 

Furthermore, as a second mode of evaluation, we also calculate performance metrics across all checkpoints to compute the average performance of the model when we do not have a parallel validation dataset to pick the best checkpoint.

As a baseline, we use the augmentation technique of aggregating noise and adding it to the clean speech, followed by passing it through a codec (g726\cite{gaoresearchG726Speech}) to introduce distortions.

Generated samples from experiments can be heard at this website
\footnote{https://github.com/leandermaben/GANSpeechAugment}. The codebase for training and testing the GANs can be found at this online repository\footnote{https://leandermaben.github.io/GANSpeechAugment/}.

\subsection{Metrics Used}
To measure model performance, we use the metrics Log Spectral Distance (LSD) \cite{shirol} and Multi-Scale Spectral Loss (MSSL)\cite{engel2020ddsp}. Both of these metrics measure the distance or dissimilarity between the generated output and the target output, which implies that the lower the value of these metrics, the better the quality of the generated output.

\subsection{Result Analysis}

The metrics computed for the test set using the GAN model checkpoints with the best validation MSSL are shown in Table \ref{tab:valid_results}.
The average of metrics for the validation set computed across all the checkpoints of the GANs is shown in Table \ref{tab:avg_results}.
Sample spectrograms from the experiments are depicted in figure \ref{fig:samples}
\begin{table}
    \centering
    \caption{Results using the validation set to pick the best checkpoint. LSD stands for Log Spectral Distance and MSSL stands for Multi-Scale Spectral Loss. Cabin and Helicopter correspond to noise types from the TIMIT dataset.}
    \resizebox{.4\textwidth}{!}{
    \begin{tabular}{|c|c|c|c|c|}
    \hline
         Dataset & Model & Mean LSD & Mean MSSL & Mode \\
         \hline
         \multirow{5}{*}{\centering RATS}& SpeechAttentionGAN& 7.29& 5.98 &Non Parallel \\
         \cline{2-5}
         &MaskCycleGAN-Augment& 7.24 & 6.10 & Non Parallel\\
         \cline{2-5}
         &SimuGAN& \textbf{6.23} & \textbf{5.26}& Non Parallel \\
         \cline{2-5}
         &Speech2Speech-Augment& 7.02 & 5.83 &Parallel\\
         \cline{2-5}
         &Baseline& 12.29 & 11.91 & -\\
         \hline
         \multirow{5}{*}{\centering Cabin}& SpeechAttentionGAN& 6.35 & 4.92 & Non Parallel\\
         \cline{2-5}
         &MaskCycleGAN-Augment& 7.56 & 6.12 & Non Parallel \\
         \cline{2-5}
         &SimuGAN& \textbf{5.48} & 4.84 & Non Parallel \\
         \cline{2-5}
         &Speech2Speech-Augment& 5.95 & \textbf{4.69} & Parallel\\
         \cline{2-5}
         &Baseline& 7.70 & 6.60 & -\\
         \hline
         \multirow{5}{*}{\centering Helicopter}& SpeechAttentionGAN& 7.16 & \textbf{5.32} & Non Parallel \\
         \cline{2-5}
         &MaskCycleGAN-Augment& 8.50 & 6.48 & Non Parallel \\
         \cline{2-5}
         &SimuGAN& \textbf{6.98} & 5.61 & Non Parallel\\
         \cline{2-5}
         &Speech2Speech-Augment& 7.17 & 5.39 & Parallel\\
         \cline{2-5}
         &Baseline& 8.53 & 6.90& -\\
         \hline
         \multirow{4}{*}{\centering Codec2}& SpeechAttentionGAN& \textbf{6.18} & 6.78 & Non Parallel \\
         \cline{2-5}
         &MaskCycleGAN-Augment& 8.09 & 7.54 & Non Parallel\\
         \cline{2-5}
         &SimuGAN& 6.28 & 7.04 & Non Parallel\\
         \cline{2-5}
         &Speech2Speech-Augment& 6.55 & \textbf{6.58} & Parallel\\
         \hline
    \end{tabular}}
    
    \label{tab:valid_results}
\end{table}

\subsubsection{RATS Dataset (UHF/VHF)}

For the RATS dataset, we find that SimuGAN outperforms the other models when we pick the best checkpoint using a validation set. It outperforms the baseline by 49.3\% and 55.8\% with respect to LSD and MSSL, respectively. However, on computing the average of metrics across all checkpoints, we find that Speech2Speech-Augment has the best results with an improvement of 43.5\% and 50.5\% as compared to the baseline in terms of LSD and MSSL, respectively.

\subsubsection{TIMIT Cabin Noise (Stationary Additive)}

In Table \ref{tab:valid_results}, in terms of LSD, SimuGAN gives the best performance with an improvement of 28.8\% as compared to the baseline. Speech2Speech-Augment performs the best in terms of MSSL with an improvement of 28.9\% with respect to the baseline. Using the second mode of evaluation as shown in Table \ref{tab:avg_results}, we find that Speech2Speech-Augment gives the best performance in terms of both LSD and MSSL with improvements of 17.1\% and 21.2\% respectively, as compared to the baseline.

\subsubsection{TIMIT Helicopter Noise (Non-Stationary Additive)}
Using the validation set, we find that SimuGAN outperforms other techniques in terms of LSD with an improvement of 18.1\% over the baseline. Using the same evaluation mode, SpeechAttentionGAN performs the best in terms of MSSL, with an improvement of 22.8\% with respect to the baseline. Using the average across checkpoints, we find that Speech2Speech-Augment outperforms the other models in terms of both LSD and MSSL with improvements of 11.7\% and 16.9\%, respectively, over the baseline.

\subsubsection{Codec2}
For Codec2 distortions, we find that SpeechAttentionGAN gives the most desirable LSD in both evaluation modes. Speech2Speech-Augment gives the most desirable MSSL in both modes.

\subsection{Limitations}

This work has two main limitations that can be explored and worked upon in future research. Firstly, we do not study how the performance of GANs varies across different languages, accent variations, age groups, and genders. Secondly, we only generate the magnitude spectrogram using GANs and recycle the phase spectrogram from the clean speech. Using GANs or other models to simulate the phase spectrogram would be a promising exploration avenue to improve simulation performance further.

\begin{table}
    \centering
    \caption{Average metrics over all checkpoints. LSD stands for Log Spectral Distance, and MSSL stands for Multi-Scale Spectral Loss. Cabin and Helicopter correspond to noise types from the TIMIT dataset.}
    \resizebox{.4\textwidth}{!}{
    \begin{tabular}{|c|c|c|c|c|}
    \hline
         Dataset & Model & Mean LSD & Mean MSSL & Mode \\
         \hline
         \multirow{5}{*}{\centering RATS}& SpeechAttentionGAN& 7.51 & 6.24 &Non Parallel \\
         \cline{2-5}
         &MaskCycleGAN-Augment& 7.51 & 6.32 & Non Parallel\\
         \cline{2-5}
         &SimuGAN& 7.77 & 7.82& Non Parallel \\
         \cline{2-5}
         &Speech2Speech-Augment& \textbf{6.94} & \textbf{5.90} &Parallel\\
         \cline{2-5}
         &Baseline& 12.29 & 11.91 & -\\
         \hline
         \multirow{5}{*}{\centering Cabin}& SpeechAttentionGAN& 7.04 & 5.71 & Non Parallel\\
         \cline{2-5}
         &MaskCycleGAN-Augment& 7.91 & 6.61 & Non Parallel \\
         \cline{2-5}
         &SimuGAN& 6.61 & 8.09 & Non Parallel \\
         \cline{2-5}
         &Speech2Speech-Augment& \textbf{6.38} & \textbf{5.20} & Parallel\\
         \cline{2-5}
         &Baseline& 7.70 & 6.60 & -\\
         \hline
         \multirow{5}{*}{\centering Helicopter}& SpeechAttentionGAN& 7.86 & 5.89 & Non Parallel \\
         \cline{2-5}
         &MaskCycleGAN-Augment& 9.01 & 7.15 & Non Parallel \\
         \cline{2-5}
         &SimuGAN& 8.32 & 7.20 & Non Parallel\\
         \cline{2-5}
         &Speech2Speech-Augment& \textbf{7.53} & \textbf{5.73} & Parallel\\
         \cline{2-5}
         &Baseline& 8.53 & 6.90& -\\
         \hline
         \multirow{4}{*}{\centering Codec2}& SpeechAttentionGAN& \textbf{6.05} & 7.11 & Non Parallel \\
         \cline{2-5}
         &MaskCycleGAN-Augment& 7.64 & 8.00 & Non Parallel\\
         \cline{2-5}
         &SimuGAN& 6.96 & 9.14 & Non Parallel\\
         \cline{2-5}
         &Speech2Speech-Augment& 6.55 & \textbf{6.55} & Parallel\\
         \hline
    \end{tabular}}
    
    \label{tab:avg_results}
\end{table}

\section{Conclusions}

We perform a detailed study of the use of GANs in simulating different varieties of noises, including UHF/VHF, Additive Stationary and Non-Stationary, and Codec distortion. Through the study, we demonstrate that different types of GANs can be successfully used for noise simulation and that the performance of these GANs varies depending on the noise types. In addition, we show that GANs can be used for noise simulation even with small training datasets of the magnitude of 3 minutes. We also find that GANs give significant improvements over the baseline technique of using aggregated noise followed by codec (g726) distortion. We explore both parallel and non-parallel translation to evaluate the capacity of GANs in both modes of translation. These results would, in turn, help train more robust models for downstream processes. Finally, we document the performance of the various GANs on different datasets in terms of LSD and MSSL, which would help provide a framework to pick the right type of GAN for the desired task.

\bibliographystyle{IEEEtran}
\bibliography{template}

\end{document}